\pdfoutput=1
\documentclass[letterpaper, 10 pt, conference]{ieeeconf}  
\IEEEoverridecommandlockouts                              
\usepackage[hyperindex,breaklinks]{hyperref}

\title{\LARGE \bf
Requirements for storing electrophysiology data
}

\author{Jeff Teeters$^{1*}$, Jan Benda$^{2}$, Andrew Davison$^{3}$, Richard C. Gerkin$^{4}$, Jeffrey Grethe$^{5}$,Jan Grewe$^{2}$,\\
Kenneth Harris$^{6}$, Christian Kellner$^{7}$, Yann Le Franc$^{8}$, Roman Mou{\v c}ek$^{9}$, Dimiter Prodanov$^{10}$,\\
Robert Pr{\" o}pper$^{11}$, Hyrum L. Sessions$^{12}$, Leslie Smith$^{13}$, Andrey Sobolev$^{7}$, Friedrich Sommer$^{1**}$,\\
Adrian Stoewer$^{7}$, Thomas Wachtler$^{7**}$, Barry Wark$^{14}$
\thanks{This work was supported by the International Neuroinformatics Coordinating Facility (INCF)}
\thanks{$^{*}$Working group leader}%
\thanks{$^{**}$Co-leader of task force}%
\thanks{$^{1}$University of California, Berkeley, USA}%
\thanks{$^{2}$Eberhard Karls Universit{\" a}t T{\" u}bingen, T{\" u}bingen , Germany}%
\thanks{$^{3}$CNRS, Gif-sur-Yvette, France}%
\thanks{$^{4}$Arizona State University, USA}%
\thanks{$^{5}$University of California, San Diego CA, USA}%
\thanks{$^{6}$Imperial College London, United Kingdom}%
\thanks{$^{7}$Ludwig-Maximilians-Universit{\"a}t M{\"u}nchen, Germany}%
\thanks{$^{8}$University of Antwerp, Belgium}%
\thanks{$^{9}$University of West Bohemia, Pilsen, Czech Republic}%
\thanks{$^{10}$IMEC, Leuven, Belgium}%
\thanks{$^{11}$Technical University of Berlin, Germany}%
\thanks{$^{12}$Blackrock Microsystems, Salt Lake City UT, USA}%
\thanks{$^{13}$University of Stirling, Scotland, UK}%
\thanks{$^{14}$Physion, Boston, USA}%
}

\begin{document}

\maketitle
\thispagestyle{empty}
\pagestyle{empty}

\begin{abstract}

The purpose of this document is to specify the basic data types required for storing electrophysiology and optical imaging data to facilitate computer-based neuroscience studies and data sharing.  
These requirements are being developed within a working group of the Electrophysiology Task Force in the International Neuroinformatics Coordinating Facility (INCF) Program on Standards for Data Sharing.  
While this document describes the requirements of the standard independent of the actual storage technology, the Task Force has recommended basing a standard on HDF5\cite{hdf5}. 
This is in line with a number of groups who are already using HDF5 to store electrophysiology data, although currently without being based on a standard. 

\end{abstract}

\medskip
\section{\textbf{Introduction}}

\subsection{Related existing systems}
This document summarizes input from task force members based on their experience on working with electrophysiology data. 
It also takes into account conventions from previous systems developed for storing electrophysiology data.  
Perhaps the most influential previous system is Neuroshare\cite{neuroshare}, which defines an API for reading data from different file formats created by manufacturers of recording equipment. 
Neuroshare defines four basic entities, i.e. data types that have been included in most subsequently developed systems.  
NEO\cite{neo} defines Python objects which can be used to represent electrophysiology data and allows interfacing to different storage backends, including Neuroshare (for reading data) and HDF5\cite{hdf5}.  
The CARMEN NDF\cite{carmen} format prescribes a method for storing electrophysiology data in MATLAB file format.  
The Kwik\cite{kwik} format used by the KlustaKwik spike sorting suite is an HDF5-based system designed for high-channel count electrophysiology data. 
The MIEN\cite{mien} system has some entities not included in other types, such as a histogram, and also accommodates cell morphology data. 
Our requirements adopt conventions from these existing systems, whenever adequate.%

In addition, there are other systems and conventions for storing  electrophysiology data that are relevant for our requirements, including: PhysioBank WFDB - WaveForm DataBase\cite{wfdb}, MEF - Multiscale Electrophysiology File Format\cite{mef}, EDF - European Data Format\cite{edf}, GDF - General data format for biosignals\cite{gdf}, the G-Node Data API\cite{g-node}, the NIX format\cite{nix}, the BRAINformat framework\cite{brainformat}, the BrainVision data format\cite{brainvision}, and the EEG/ERP Portal (EEGBase)\cite{eegbase}. 
The most recent is Neurodata Without Borders (NWB)\cite{nwb}, which has already been used by the Allen Institute for Brain Science to share a large volume of slice physiology data.

Systems for storing neuroimaging data should also be considered.  
A few are: MINC\cite{minc} (which uses NetCDF and HDF5), and Nifti\cite{nifti}.  
For an HDF5 implementation, methods used for NSDF - Neuroscience Simulation Data Format\cite{nsdf}, MTSF\cite{pfeiffer}, shared retina data format\cite{eglen}, and NeXus format could be relevant\cite{nexus}.

\medskip
\section{\textbf{Scope of this Document}}
The document concerns the requirements for the storing of electrophysiology data and cellular optical imaging data, that is, measurements of neuronal activity over time.  
To make data useful it is necessary to provide metadata describing how to interpret the stored numbers and to document the entire experimental context. 
The scope of this specification is to specify what needs to be stored to account for the most commonly obtained cellular-level neurophysiology data and for the minimal metadata required to interpret the stored numbers.  
This includes data from the following experimental modalities, \textit{in vitro} or \textit{in vivo}:

\begin{itemize}
\item Patch clamp or sharp electrodes recordings
\item Single or multi-electrode probes
\item Multielectrode arrays (MEA)
\item EEG/ERP
\item ECOG
\item MEG
\item Optical imaging of cellular activity
\end{itemize}

Although cellular-level neurophysiology data is the main focus of this document, other types of measurements acquired during an experiment may be included as well.  
For example, movements, blood oxygenation level, or heart rate.  
It is desirable that the standard be general enough to allow including additional data obtained from an experiment.

\subsection{What is not in this document}
Data sets often include results of many sessions and stimuli stored in multiple files.  
This initial standard does not specify the metadata required for understanding the overall context or the multi-file structure of data sets, for example:

\begin{itemize}
\item Metadata common to the entire data set, such as purpose of experiments, lab, and publications.
\item Inventory of all data files constituting a data set.
\item Specification of stimuli.
\end{itemize}

However, provisions should be made to accommodate the storage of this data.  

\subsection{Definitions}
The following definitions apply when the terms are used.

\begin{itemize}
\item Time: Date, local time, and time zone.  
\item REQUIRED, MUST, SHOULD, and MAY are used in accordance with RFC 2119. In particular, REQUIRED/MUST denote a requirement of a compliant implementation that infomation must be stored with sufficient precision to be retrieved accurately.
\item CHALLENGES: Use cases that the standard must be able to accommodate.
\end{itemize}

\medskip
\section{\textbf{Global Metadata}}
For each data file there MUST contain metadata that make it possible to obtain:

\begin{itemize}
\item Version identification of the  file format.
\item Identification metadata for describing the origin of the data file.   
Possible information for this includes lab, experimenters’ names.  
The contents and format is not specified in this document.
\item A unique id (UUID, for example) that can be referenced if the data set is shared. 
\end{itemize}

\medskip
\section{\textbf{Requirements for each data type}}
An overview of the data types is given in Table 1.
\begin{table*}[t]
\begin{tabular}{ c | p{14.3cm} }
  \hline			
  \textbf{Data type} & \textbf{Description}\\
  \hline
  Signal source & The origin of the recorded data; for example, the identity, position, etc. of the recording implement.  Must support a hierarchy, for example a tetrode source contains multiple electrode sources. \\
  Time series & An ordered collection of values given at defined points in time; for example, a recorded voltage signal.  Must support regularly and irregularly sampled values. \\
  Signal events & Events defined by the content of a signal that occur at defined times; for example, spikes, synaptic potentials, artifacts.  Also includes properties associated with signal events.  For example, spike waveforms, features used for spike sorting, classification of putative spikes. \\
  Image stacks & An ordered collection of images; meaning of stack dimensions must be specified.  For example, a z-stack of images, a time series of images, a time-series of z-stacks. \\
  Experimental events & Events defined by the design of the experiment, that occur at specified points or regions in time; for example, a stimulus, or an animal behavior. \\
  Other / Generic array & Other data types and a mechanism for storing data types which have not been included in the standard. \\
  \hline  
\end{tabular}
\caption{Overview of data types required}
\end{table*}

\subsection{Signal source}
The origin of the recorded data; for example, the identity, position, etc. of the recording. This typically refers to information about a channel name, description, and location, or the target recorded.  There are at least two kinds of sources, including actors doing the recording (e.g. an electrode) and objects/actors being recorded (e.g. an subject, brain region, or neuron (real or putative))
\smallskip
\subsubsection*{REQUIRED}
\begin{itemize}
\item Type of source.  The following types must be accommodated:
 \begin{itemize}
 \item Recording equipment (doing the recording):
  \begin{itemize}
  \item Electrode
  \item Electrode array (e.g. tetrode, shank, Utah Array, ...).
  \item Amplifier
  \end{itemize}
 \item Biological source (things being recorded):
  \begin{itemize}
  \item Subject
  \item Brain region
  \item Unsorted MUA (Multiunit activity).
  \item Neuron (intracellular or identified by sorting extracellular signals)
  \end{itemize}
 \end{itemize}
\item An identifier for a source, e.g. a name or numeric ID.
\item Must support hierarchical sources, for example a tetrode source contains multiple electrode sources, or a brain region contains multiple neurons.  
\item Metadata about the source, e.g. the model of the recording instrument or the location of the neuron recorded
\end{itemize}
\smallskip
\subsubsection*{CHALLENGES}
\begin{itemize}
\item Recording channels have spatial locations whose relative positions may be important.  Recording channels may be located near each other, for example, on the same tetrode, so that a spike may be detected on more than one channel.  Also, data recorded from the same channel but at different times are related since the signal source is the same.
\item A signal source could be derived from a region of interest (ROI) in an image stack.  The position and shape of the ROI could vary over time.
\item There may be properties of a signal source that are invariant to multiple sessions (e.g. electrode location) and other properties that change with each session (e.g. amplifier gain).
\end{itemize}

\subsection{Time series}
An ordered collection of values given at defined points in time; for example, a recorded voltage  signal. It may correspond to raw data, recorded from “electrodes” or “channels”.   Alternatively, they could be derived from a data processing step.  The sampling may be done at regular time intervals (the sampling rate) or at irregular intervals (in which case a time point is required for each value).
\smallskip
\smallskip
\subsubsection*{REQUIRED}
\begin{itemize}
\item The sample values, including units.  
\item The time point corresponding to each index in the data.  For example, for irregularly sampled data, the times of sampling must be represented. 
\item Starting time. 
\item Duration.  For example, in an irregularly sampled time series, the start and end time might require explicit representation.  
\item Sampling rate (for regularly sampled data).
\item For derived data, provenance information.  
\item Descriptive label.
\item Signal sources, for example channels. 
\end{itemize}
\smallskip
\subsubsection*{CHALLENGES}
\begin{itemize}
\item Sometimes samples get dropped or out of sync.  Therefore, the time point derived from the index of a regularly sampled time series may be incorrect if derived in the traditional fashion.  It is possible to correct this by storing information that reflects when these desynchronizing events occurred, or reference time points for resynchronization.  In general, redundant mechanisms for ensuring the accuracy of time points are encouraged.  
\item Multiple time series may have unrelated sampling rates. 
\end{itemize}

\subsection{Signal events}
The Signal events identify changes in a signal; for example, spikes, synaptic potentials (PSPs), and artifacts.  They could be raw data, for example, output of a hardware device that detects, and provides the times of spikes. They may also be generated as the result of processing data, for example, applying a spike detection algorithm on a time series signal.
\smallskip
\smallskip
\subsubsection*{REQUIRED}
\begin{itemize}
\item Starting time of recording
\item Duration of recording
\item The time of each event
\item Properties common to all events in the session.  It must be possible to store the following properties:
 \begin{itemize}
 \item The source channel(s) the events were derived from (if the same for all events)
 \item Spike templates used to do spike sorting
 \item Description of how neural events were detected in the signal
 \item Trigger information, i.e. the trigger type and threshold of trigger used to detect spikes.  
 \end{itemize}
\item Properties associated with each event. (The Signal event properties).  These include:
 \begin{itemize}
 \item The source channel(s) the event was derived from (if varies between events).
 \item Spike waveform and metadata needed to interpret the waveform (sampling rate, units of measure), including timing information relative to source signal.
 \item Identity of neuron(s), associated with the spike event, often resulting from spike sorting.  The following types of unit assignments MUST be accommodated:
  \begin{itemize}
  \item At most only one unit associated with each event.
  \item Possible to have more than one unit associated with each event.
  \item A probabilistic association of multiple units with each event.
  \end{itemize}
 \item Feature vectors used for spike sorting.
 \end{itemize}
\end{itemize}
\smallskip
\subsubsection*{CHALLENGES}
\begin{itemize}
\item If there are multiple channels and the trigger information varies between channels, the trigger information for each channel must be specified.
\item It should be possible to associate multiple sets of properties to the same Signal events.  For example, multiple feature vectors and multiple neuron identities result from different spike sorting methods.
\item Event waveforms lengths may be variable.
\end{itemize}

\subsection{Image stacks}
Image stacks are an ordered collection of images; the meaning of stack dimensions must be specified.  For example, a z-stack of images, a time series of images, a time-series of z-stacks, movies which are in standard formats (mpeg, mp4).
\smallskip
\subsubsection*{REQUIRED}
\begin{itemize}
\item Starting time
\item Duration
\item The collection of images making up the image stack.
\item Spatial relationship between sources contributing to each image; both within an image and between different images recorded at the same time (this could be specified in the Sources that are associated with each image).
\item Time that each image is recorded.
\item Units of measure of each image pixel (e.g. RGB, gray scale, voltages).
\end{itemize}
\smallskip
\subsubsection*{CHALLENGES}
\begin{itemize}
\item The sources which contribute to each image pixel may not be organized in a rectangular grid, but could be in a different shaped grid (such as hexagonal) or irregularly spaced.
\end{itemize}

\subsection{Experimental events}
The Experimental Events data type consists of times of events, along with values that correspond to the times.   This data type can be used to describe: stimuli, trials or sweeps, time intervals, behaviors, and other events or conditions that occur during the experiment.
\smallskip
\subsubsection*{REQUIRED}
\begin{itemize}
\item Starting time of event monitoring (earliest time an event could be detected)
\item Ending time of event monitoring (latest time that an event could be detected)
\item Description of the events
\item Sequence of event times
\item Property values needed to describe each event.  These could be a string or numeric value, a list of values together with metadata required for interpreting the property values.  For example, if the event were the start of presentation of a sound of fixed frequency, the metadata would specify that the numeric values indicate the frequency in Hz.
\end{itemize}

\subsection{Other / generic array}
This section describes types of data that must be included in a standard which have not been specified above.  For example, a generic array type may be useful to store arbitrary data, such as a histogram of event times or spike counts.  It may also be desirable to include a more general “Data aggregation” type in a format. 
\smallskip
\subsubsection*{REQUIRED}
\begin{itemize}
\item Textual description of the array contents.
\item List of categories, for categorical data aggregations.
 \begin{itemize}
 \item For each category, count of instances or relative quantity associated with that category.
 \end{itemize}
\item An array (of any number of dimensions). The size should be specified, e.g. N x M x T x …
Description.
\item Description of each dimension (including unit, if applicable).
\item Heading and description of the contents of each slice (e.g. column), if applicable.
\item Reference to any entities (Source channels, Image Stacks, Neural Events ...) containing related data and the nature of the relationship.
\end{itemize}

\medskip
\section{\textbf{Relationships}}
Relationships are metadata for describing associations between different entries in a data set.  At least the following types of relationships MUST be expressible: 

\begin{enumerate}
\item \textit{Derived from}.  Data that is derived from other data is related to the data it is derived from. In general, it is a transformation or process that links inputs (for example, segments of collected data) to outputs (for example, features or derived data).  For example, the relationship between spike waveform data in its raw form and its representation in a new basis (e.g. PCA) must be stored; or signal event data can be related to time series data and/or segment data (Spike waveforms) from which it was extracted.  

For every \textit{derived from} relationship, it MUST be possible to store information describing how the derivation was performed.  To accomplish this, use of the W3C Prov Data Model, is recommended (OPTIONAL). In Prov terminology, the Entities are defined by the ends of the relationship, so at minimum, it MUST be possible to store Agent(s) and Activities associated with the derivation. It SHOULD be possible to inspect and search this provenance information.

\item \textit{Grouping}.  There MUST be a way to extract a logical grouping of experiment data, as intended by the design of the experiment: for example, several experimental sessions of recording from a single subject, or several trials from the same session.  
A conclusion section is not required. Although a conclusion may review the main points of the paper, do not replicate the abstract as the conclusion. A conclusion might elaborate on the importance of the work or suggest applications and extensions. 
\end{enumerate}

\medskip
\section{\textbf{Application Programming Interface (API)}}
An API for accessing stored electrophysiology data is REQUIRED. The API MUST:
\begin{itemize}
\item Enable storing and retrieving all data and metadata described in this document.
\item Offer the opportunity to structurally validate the contents of data files.
\end{itemize}
Any API for retrieving data SHOULD provide the following functionality:
\begin{enumerate}
\item Retrieve global metadata.
\item Retrieve an inventory/enumeration of all data entities stored in the data file.  This inventory should include: data type and dimensions; and, if applicable: starting time and duration.
\item For a given data entity, retrieve the associated metadata.
\item For a given data entity, retrieve a section or slice of the data.
\item For a given source channel, retrieve data entities recorded using the source.
\item For a given data entity, find all related data entities (referenced in the metadata).
\item Retrieve an inventory of all logical groupings.
\item For a given logical grouping, retrieve metadata about each data entity within the group.
\end{enumerate}

\addtolength{\textheight}{-12cm}   


\end{document}